\documentclass[pra,twocolumn,longbibliography,nofootinbib,floatfix]{revtex4-1}
\usepackage[utf8]{inputenc}
\usepackage{bm,amsfonts,amsmath,amssymb}
\usepackage{dcolumn,xfrac}
\usepackage{graphicx}
\usepackage{natbib}
\usepackage{braket}
\usepackage{tipa}
\usepackage{booktabs}
\usepackage{multirow}
\usepackage{cases}
\usepackage{dcolumn}
\newcommand{\Eref}[1]{Eq.~(\ref{#1})}
\newcommand{\Tref}[1]{Table~\ref{#1}}
\newcommand{\Fref}[1]{Figure~\ref{#1}}
\def\andname{\hspace*{-0.5em}}

\usepackage{mathtools}

\begin{document}
\title{The Bohr-Weisskopf effect in the potassium isotopes}
\author{Yu. A. Demidov$^{1,2,3}$}
\author{M. G. Kozlov$^{1,2}$}
\author{A. E. Barzakh$^{1}$}
\author{V. A. Yerokhin$^{3}$}
\affiliation{$^1$ Petersburg Nuclear Physics Institute NRC ``Kurchatov Institute'', Gatchina 188300, Russia \\
$^2$ St.~Petersburg Electrotechnical University ``LETI'', St. Petersburg 197376, Russia\\
$^3$ Peter the Great St. Petersburg Polytechnic University, St. Petersburg 195251, Russia}
\begin{abstract}
The magnetic hyperfine structure constants have been calculated for low-lying levels in neutral potassium atom taking into account the Bohr--Weisskopf (BW) and Breit--Rosenthal (BR) effects.  According to our results the $4p_{1/2}$ state of K~I is free from both BR and BW corrections on the level of the current theoretical uncertainties.
Using this finding and the measured values of the $A(4p_{1/2})$ constants, we corrected the nuclear magnetic moments for several short-lived potassium isotopes.
The BW correction is represented as a product of atomic and nuclear factors. We calculated the atomic factor for the ground state of K I, which  allowed us to extract nuclear factors for potassium $I^\pi = 3/2^+$ isotopes from the experimental data. In this way the application range of the single-particle nuclear model for nuclear-factor calculation in these isotopes has been clarified.
\end{abstract}
\maketitle
\section{Introduction}
The magnetic-dipole hyperfine structure (HFS) constants are highly sensitive to the changes of charge and magnetization distributions inside the nucleus because these constants are defined by the behavior of the electron wave function in this region. High experimental accuracy is achieved in the spectroscopic measurement of HFS constants for atoms, which allows one to study the nuclear effects in isotope sequences. These experimental data are very useful for understanding of properties of the atomic nuclei.

The HFS constant $A$ for the finite nucleus can be written in the following form \cite{stroke61},
\begin{align}\label{HFS_param}
    A &= g_I {\cal A}_0 (1-\delta)(1-\varepsilon)\,,
\end{align}
where $g_I = \frac{\mu}{\mu_N I}$ is the nuclear $g$ factor,  $\mu$ and $I$ are magnetic moment and spin of the nucleus, respectively; $\mu_N$ is nuclear magneton. $g_I {\cal A}_0$ is the HFS constant for the point-like nucleus, $\delta$ and $\varepsilon$ are Breit--Rosenthal~\cite{RB32,CS49} (BR) and Bohr--Weisskopf \cite{BW50} (BW) corrections, respectively.

For stable or long-lived isotopes, measurements of the nuclear $g$-factor and the HFS constant can be carried out independently.
These experimental data enable one to evaluate a relative hyperfine anomaly $^1\Delta^2$ (RHFA) values through the relation,
\begin{align} \label{rhfa}
^1\Delta^2 \equiv \frac{g_I^{(2)}A^{(1)}}{g_I^{(1)}A^{(2)}}-1 &\approx \varepsilon^{(2)} - \varepsilon^{(1)} + \delta^{(2)} - \delta^{(1)}=\\ \nonumber
                                                                &= ^1\Delta^{2}_\mathrm{BW} + ^1\Delta^{2}_\mathrm{BR}.
\end{align}
Here, nuclear $g$-factors, $A$-constant values, BR and BW corrections for isotopes (1) and (2) are marked by the corresponding superscript.

The dependence of BR correction on the nuclear radius $R$ is defined by the asymptotic behaviour of the electron wavefunction near a point nucleus~\cite{ionesco60}.
Then, the BR correction for $s_{1/2}$ and $p_{1/2}$ atomic states can be written as~\cite{ionesco60,Sha94},
\begin{align}\label{BR-bN}
\delta(R) = b_N (R/\lambdabar_C)^\varkappa\,, 
    \quad \varkappa = 2\sqrt{1-(\alpha Z)^2}-1\,.
\end{align}
Here $\lambdabar_C$ is the reduced Compton wavelength of the electron ($\lambdabar_C =\tfrac{\hbar}{m_e c}$), $\alpha$ is the fine structure constant, $Z$ is nuclear charge, and dimensionless parameter $b_N$ depends on the electron state. Taking into account that the charge density is almost homogeneous inside the nucleus~\cite{fermi_model} we use \mbox{$R = \sqrt{5/3}\, r_\mathrm{rms}$}, where \mbox{$r_\mathrm{rms}=\langle r^2\rangle^{1/2}$} is a root-mean-square nuclear charge radius.

Assuming the atomic-nuclear factorization, the BW correction takes the form~\cite{KDKB18,skripn20},
\begin{align} \label{BW-bM}
    \varepsilon (d_\mathrm{nuc},\,R) = d_\mathrm{nuc}\,\varepsilon_\mathrm{at} (R), \quad \varepsilon_\mathrm{at}(R)= b_M (R/\lambdabar_C)^\varkappa.
\end{align}
The accuracy of such separation had been found to be very high~\cite{skripn20,pros21}. In the case of the point-like magnetic dipole $d_\mathrm{nuc}=0$, whereas the homogeniously magnetized sphere of radius $R$ corresponds to $d_\mathrm{nuc}=1$.

The HFS constants for $p_{3/2}$ and other electronic states with angular momentum $j\ge 3/2$ are sensitive to the nuclear charge and magnetization distributions only due to the admixture of $s_{1/2}$ and $p_{1/2}$ partial waves~(see Refs.~\cite{KKDB17,PMS19}). Therefore the BR and BW corrections for all electron states are described by Eqs.~(\ref{BR-bN}) and (\ref{BW-bM}), respectevly.

The parameterization of HFS constants by Eqs.~(\ref{HFS_param}) -- (\ref{BW-bM}) involves three nuclear ($g_I$, $d_\mathrm{nuc}$, and $R$) and three atomic (${\cal A}_0$, $b_N$, and $b_M$) characteristics. In order to perform an atomic-structure calculation of the $A$ constants we need to fix the nuclear parameters,
\begin{align}\label{HFS_nuc}
    A &= A(g_I,d_\mathrm{nuc},R)\, =\,g_IA(1,d_\mathrm{nuc},R)\, .
\end{align}
The atomic parameters are the same for different isotopes and obtained numerically. The $b_M$ parameter can be found from Eqs.~\eqref{HFS_param} and \eqref{BW-bM}:
\begin{align}\label{b_M}
	b_M &= \frac{\lambdabar_C^\varkappa}{R^\varkappa} 
    \left(1 - 
    \frac{A(g_I,1,R)}{A(g_I,0,R)}
    \right)\,.
\end{align}
To find parameter $b_N$ we performed calculations for two different nuclear radii: 
\begin{align}\label{b_N}
    b_N &= 
	\frac{\left(A\left(g_I,0,R_2\right) - A\left(g_I,0,R_1\right)\right ) \lambdabar_C^\varkappa}
    {A(g_I,0,R_2)R_1^\varkappa -A(g_I,0,R_1)R_2^\varkappa}
    \,.
\end{align}
The atomic parameter ${\cal A}_0$ was found from the relation:
\begin{align}\label{A_0}
    {\cal A}_0 &= 
    \frac{A(1,0,R)}
	{1 - b_N (R/\lambdabar_C)^{\varkappa}} 
    \,.
\end{align}

The independent measurements of the HFS constants and nuclear magnetic moments allow to determine RHFA for several K isotopes~\cite{Per13}. At the same time the electronic structure of potassium atom is relatively simple and consists of a single valence electron above the filled atomic core. Advanced atomic methods allow to calculate HFS constants of potassium isotopes with high accuracy~\cite{owusu97,saf99,saf_k}.   The BW effect should no longer be ignored at the level of accuracy modern experiments and theoretical atomic-structure calculations.

In Refs.~\cite{papuga13,papuga14} the HFS measurements for potassium isotopes were extended up to $^{51}$K, enabling one to assess the nuclear magnetic moments. In the present work we recalculate these nuclear magnetic moments taking into account hyperfine anomaly corrections.

\section{Hyperfine anomaly}
In Sec.~4 we will show that for all pottasium isotopes considered here $\Delta_\mathrm{BR}$ is three orders of magnitude smaller than $\Delta_\mathrm{BW}$.
Correspondingly, we can neglect BR contribution to RHFA and assume \mbox{$\Delta \approx\ \Delta_\mathrm{BW}$}.
Then one can determine $d_\mathrm{nuc}^{(2)}$ for the isotope in question provided the nuclear factor $d_\mathrm{nuc}^{(1)}$ for the reference isotope, RHFA value $^1\Delta^2$, and the atomic part of BW correction $\varepsilon_\mathrm{at}$ are known,
\begin{align}\label{rhfa2}
d_\mathrm{nuc}^{(2)} = d_\mathrm{nuc}^{(1)} +\frac{^1\Delta^{2}}{\varepsilon_\mathrm{at}}.
\end{align}
These factors can be compared with that calculated within a framework of the single-particle nuclear model~\cite{BW50,B51}. 
One can expect that this model works fairly well for $^{39}$K with one proton hole ($\pi d_{3/2}^{-1}$) with respect to the doubly magic $^{40}$Ca. At the same time, it was shown~\cite{stone73} that BW correction (i.e. the $d_\mathrm{nuc}$ factor) for $\pi d_{3/2}$ state is anomalously large and sensitive to small perturbations, for example, in the case of gold \mbox{$I^{\pi}=3/2^+$} isotopes~\cite{dem20,B2020,roberts21}. Thus, the study of \mbox{$I^{\pi}=3/2^+$} potassium isotopes with one hole in the closed proton shell can give additional insight in this single-particle nuclear structure.
\section{Single-particle nuclear model}
The nuclear magnetization mainly arises due to the spin polarization and the orbital motion of the valence nucleon.
The nuclear $g$-factor is given by the famous Land{\'e} formula,
\begin{align}
\label{g_S}
\begin{split}
g_I = &\left [ \frac{1}{2} -\frac{L(L+1)-3/4}{2I(I+1)} \right]g_S\\ 
 +&\left [ \frac{1}{2} +\frac{L(L+1)-3/4}{2I(I+1)} \right]g_L\,. 
\end{split}
\end{align}
Introducing $\sigma$ (the average odd-particle spin component on the direction of $\bm{I}$) in accordance with relation:
\begin{align}
\label{4sigma}
g_I = \frac{\sigma}{I} g_S + \frac{(I-\sigma)}{I}g_L, 
\end{align}
we obtain,
\begin{subnumcases}
{\label{g_S2}\sigma =}
 \label{g_S21}
~\frac12, & \text{$I = L +\tfrac12$}\\
 \label{g_S22}
-\frac{I}{2(I+1)}, & \text{$I = L -\tfrac12$}.
\end{subnumcases}
The spin $g$-factor $g_S$ is chosen from the condition that Eqs. (\ref{g_S}) and (\ref{4sigma}) reproduce the experimental $g$-factor value by setting $g_L = 1$ for proton and $g_L = 0$ for neutron~\cite{Sha94}. Such a choice of $g_L$ gives $g_S$ within the range from 0.84$g_p^\mathrm{free}$ to 0.95$g_p^\mathrm{free}$ (the free-proton $g$ factor $g_p^\mathrm{free} = 5.586$) for the considered potassium $I^\pi = 3/2^+$ isotopes.

Then the BW correction $\varepsilon$ can be represented as a linear combination of the spin and orbital contributions, $\varepsilon_S$ and $\varepsilon_L$ with the weights determined by \Eref{4sigma},
\begin{align}
\label{eps_frac}
\begin{split}
\varepsilon = &\frac{\sigma g_S}{Ig_I}\varepsilon_S + \left(\frac{I-\sigma}{I} \right )\frac{g_L}{g_I}\varepsilon_L\,.
\end{split}
\end{align}

One can represent $\varepsilon_S$ and $\varepsilon_L$ according to \citet{BW50} in the following form,
\begin{align} 
\label{eps_SL}
\varepsilon_S = (1 - k\zeta)\varepsilon_\mathrm{at}, \quad \varepsilon_L = (1 + k)\varepsilon_\mathrm{at}\,.
\end{align}
Here, $k \approx -0.38$~\cite{BW50}, $\zeta$ is the so-called spin asymmetry parameter~\cite{bellac,B51}. If the valence nucleon is in the $L\ne 0$ state, then the spin density is asymmetric and additional contribution to the spin part of BW correction appears. Expressions for $\zeta$ were suggested by ~\citet{B51}:
\begin{subnumcases}
{\zeta = \label{zeta_both}}
\label{BW_zeta+}
\frac{2I-1}{4(I+1)}, &\text{$I = L+\tfrac12$}\\
\label{BW_zeta-}
\frac{2I+3}{4I},     &\text{$I = L-\tfrac12$.}
\end{subnumcases}
The nuclear factor can be found from Eqs.~(\ref{4sigma}) -- (\ref{eps_SL}) as,
\begin{align} 
\label{F2}
d_\mathrm{nuc} = 1 + k\left [ 1 - (1+\zeta)\frac{\sigma g_S}{Ig_I}\right ]\,.
\end{align}
\begin{table}[h!]
\caption{\label{tbl:energies}
The binding energies (in au) of the low-lying electron states of potassium atom relative to the $\rm K^+$ core.
The rows DHF, MBPT, and LCC correspond to the Dirac--Hartree--Fock, Dirac--Hartree--Fock plus MBPT, and Dirac--Hartree--Fock plus LCC methods, respectively. We take into account Breit corrections at DHF stage of the calculations. The experimental data and the theoretical error (in \%) are listed in the last two rows.}
\begin{tabular}{lcccc}
\hline
\\[-3mm]
Method                  &$4s_{1/2}$&$4p_{1/2}$& $4p_{3/2}$\\
\hline
DHF                  &{~~0.1475~~}&{~~0.0957~~}&{~~0.0955~~}\\
MBPT                 &{~~0.1609~~}&{~~0.1007~~}&{~~0.1004~~}\\
LCC                  &{~~0.1601~~}&{~~0.1006~~}&{~~0.1003~~}\\
Expt~\cite{NIST}     &{~~0.1595~~}&{~~0.1004~~}&{~~0.1001~~}\\
Diff.with expt.      &\multicolumn{1}{c}{$0.36$\%} &\multicolumn{1}{c}{$0.21$\%}&\multicolumn{1}{c}{$0.21$\%}\\
\hline 
\end{tabular}
\end{table}
When the nuclear factor is large, a more accurate estimate of the parameter $k$ given by~\Eref{eps_SL} than that of ~\cite{BW50} is needed. This parameter can be calculated directly by solving the Schr{\"o}dinger equation with the Woods-Saxon potential~\cite{nuc_so} for the valence nucleon. After that the radial wave function of the valence nucleon is used to compute the ratio $\varepsilon_L/\varepsilon_\mathrm{at}$ as proposed in Refs.~\cite{zherebtsov00,yerokhin08}. The ratio $\varepsilon_L/\varepsilon_\mathrm{at} = 0.621(2)$, corresponding to the parameter \mbox{$k = -0.379(2)$}, is quite stable for all considered potassium isotopes. Deviations between the numerical results determine the uncertainty of $k$. 
\section{ Calculation results}
We consider the ground and valence-excited configurations of the potassium atom, which can be represented as as a single valence electron above the $3s^{2}3p^{6}$ electron shells included in the atomic core. The core-valence and core-core correlations are treated perturbatively. All calculations are performed using Dirac-Coulomb-Breit Hamiltonian. Breit corrections including both the magnetic term and the retardation term in the zero-frequency limit are taken into account in accordance with Refs.~\cite{breit1,breit2}. 

We start by solving Dirac-Hartree-Fock (DHF) equations for the core and valence orbitals up to $5p_{3/2}$. 
After that we merge these orbitals with B-splines of order 8 as described in Ref.~\cite{basis} to form a basis set for calculating the correlation corrections. The basis set $22spdfgh$ includes 230 orbitals for partial waves with orbital angular momentum $l$ from 0 to 5.

Correlation corrections to the HFS include ones to the hyperfine operator and to the many-electron wave functions. To account for the correlation corrections to the effective hyperfine operator we use the random phase approximation (RPA) with structural radiation correction \cite{KPJ01}. These corrections include in particular the spin polarization of the core shells, down to $1s$. We use second order many-body perturbation theory (MBPT) \cite{DFK96b,KPST15} and linearized single double coupled-clusters method (LCC) \cite{blundell89,blundell91,SKJJ09} to take into account correlation corrections to the wave function. In both cases these corrections are included in self-energy contribution to the effective Hamiltonian for a single valence electron~\cite{DFKP98}. The energy dependence of the effective Hamiltonian is taken into account as discussed in Refs.~\cite{DFK96b,SKJJ09}. As seen from Tables 1 and 2, the LCC results agree with the experimental data better than the MBPT ones. We have already seen the same preference for LCC results in our previous calculations~\cite{dem20}.
\begin{table}[tbh]
\caption{\label{tbl:hfs_K} The atomic parameters ${\cal A}_0$, $b_N$, $b_M$, and HFS constants for the lower levels of K I. We compare HFS constants for $^{39}\mathrm{K}$ ($g_I = 0.2609775 (2)$~\cite{39K_mu} and single-particle factor $d_\mathrm{nuc} = -2.1$) with avalable experimental data~\cite{K_st_s,K_st_p}.}
\begin{tabular}{lrrrr}
\hline
\\[-3mm]
Method &${\cal A}_0$ (MHz)&$b_N$&$b_M$&$A$  (MHz)\\
\hline
\multicolumn{5}{c}{$4s_{1/2}$}\\
DHF                  &564.4&{0.218~~}&{0.079~~}   & 147.2\\
RPA                  &697.9&{0.216~~}&{0.078~~}   & 182.0\\
RPA+MBPT             &915.5&{0.206~~}&{0.078~~}   & 238.8\\
RPA+LCC              &888.1&{0.206~~}&{0.078~~}   & 231.6\\
Experiment ($^{39}\mathrm{K}$)&&\multicolumn{3}{r}{230.8598601(3)}\\
Relative error       &&&&             $0.3\%$\\
\multicolumn{5}{c}{$4p_{1/2}$}\\
DHF                  &63.6 &{0.002~~}&{0.001~~}   & 16.6\\
RPA                  &82.4 &{$-0.010$~~}&{$-0.003$~~}   & 21.5\\
RPA+MBPT             &110.1&{$-0.004$~~}&{$-0.001$~~}   & 28.7\\
RPA+LCC              &107.7 &{$-0.004$~~}&{$-0.001$~~}  & 28.1\\
Experiment ($^{39}\mathrm{K}$)&&&\multicolumn{2}{r}{ 27.775(42)}\\
Relative error       &&&&             $1.2\%$\\
\multicolumn{5}{c}{$4p_{3/2}$}\\
DHF                  &12.4  &{0.000~~}&{0.000~~}& 3.2\\
RPA                  &20.7  &{0.050~~}&{0.016~~}& 5.4\\
RPA+MBPT             &24.0  &{0.029~~}&{0.008~~}& 6.3\\
RPA+LCC              &23.4  &{0.030~~}&{0.008~~}& 6.1\\
Experiment ($^{39}\mathrm{K}$)&&&\multicolumn{2}{r}{6.093(25)}\\
Relative error       &&&&             $1.0\%$\\
\hline
\end{tabular}
\end{table}

A comparison of theoretical binding energies of the $4s_{1/2}$, $4p_{1/2}$, and $4p_{3/2}$ states with experiment is given in \Tref{tbl:energies}. The experimental data used in this comparison are from Ref.~\cite{NIST}. Our final theoretical uncertainty ranges from 130 cm$^{-1}$ for 4s to 50 cm$^{-1}$ for 4p states of K I. The theoretical result for the fine-structure $4p_{1/2} - 4p_{3/2}$ interval of 58.2 cm$^{-1}$ is in excellent agreement with experimental value, 57.7~cm$^{-1}$~\cite{NIST}.

The calculated HFS atomic parameters for $4s_{1/2}$, $4p_{1/2}$, and $4p_{3/2}$ states of potassium are given in \Tref{tbl:hfs_K}. The parameter ${\cal A}_0$ is highly sensitive to the electronic correlations treatment. The uncertainty of ${\cal A}_0$ calculations can be reliably estimated for the $4p$ states. 
The changes in $A(4p_{1/2})$ and $A(4p_{3/2})$ constants due to BR corrections are only 0.005\% and 0.04\%, respectively.
The contributions of BW corrections are of the same order of magnitude. Both BR and BW corrections can be neglected for these states in present consideration.
Thus, the deviation of the theoretical $A(4p)$ constants from the experimental values stems exclusively from the  incompleteness of ${\cal A}_0$ calculations.
Our LCC results agree with experimental data for $\rm ^{39}K$ within 1.2\% for $A(4p_{1/2})$ constant and 1.0\% for $A(4p_{3/2})$ one. 
It should be noted that taking into account partial triple excitations within LCC method significantly reduces calculation uncertainty of the $A(4p_{1/2})$ constant~\cite{saf_k}. We conservatively estimate the possible uncertainty of the ${\cal A}_0\, (4s_{1/2} )$ calculation for K I within LCC method as 1.2\%. Relative correlation contributions in ${\cal A}_0$ for $4s_{1/2}$ and $4p_{1/2}$ states are close to each other ($\sim 60$\%, see \Tref{tbl:hfs_K}), therefore, one can expect that the accuracy of the ground-state calculation is not worse than that for excited state.

The calculation of the parameter $b_N$ requires a variation of the nuclear radius, which leads to a change in the integration grid within the framework of our software package~\cite{KPST15}. Therefore, the parameter $b_N$ is more sensitive then $b_M$ to the size of the basis set. As a final $b_N$ value for $4s_{1/2}$ state in potassium we adopted the LCC result with the uncertainty covering the deviations of the results obtained in the frameworks of the different approximations (see column 3 of \Tref{tbl:hfs_K}): $b_N (4s_{1/2}) = 0.206(12)$. Then the BR correction for the ground state of $\rm ^{39}K$ is 0.26(2)\%. Using the nuclear radii from Ref.~\cite{angeli13} we found that $\prescript{39}{}\Delta^{47}_\mathrm{BR} = 4\cdot 10^{-6}$.

The $b_M$ parameter for the $4s_{1/2}$ state of potassium is stable at the each stage of correlation effects treatment. Conservatively assuming the same relative error for parameters $b_M$ and $b_N$ one can obtain $\varepsilon_\mathrm{at} = 0.098(4)\%$ for the ground state of K I.
The atomic part of BW correction is weakly dependent on the principal number of electron state~\cite{shab01}. Because of that the $\varepsilon_\mathrm{at}$ corrections calculated for $s$ states of H-like ion and neutral atom should be comparable. Our result coincides with $\varepsilon_\mathrm{at} = 0.098\%$ obtained for the ground state of H-like potassium ion~\cite{Sha94}. Note, following~\citet{B51} in Refs.~\cite{papuga13,papuga14} was used overestimated value $\varepsilon_\mathrm{at} = 0.125\%$. A comparison of atomic parameters for HFS constants of H-like ions calculated by us with results of \citet{Sha94} and \citet{B51} is given in Ref.~\cite{ita20}. 
\begin{figure}[tbh]
\includegraphics[height=8.5cm]{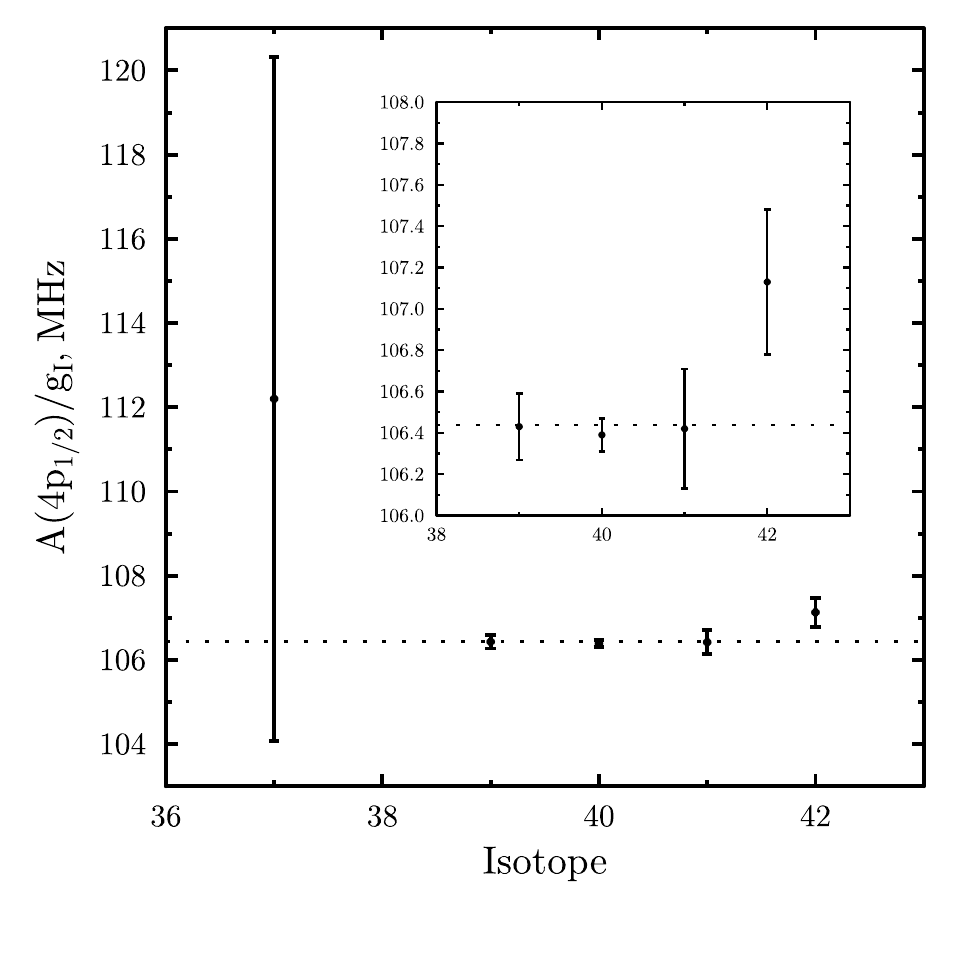}
\caption{\label{exp} \textbf{FIG.~1.} The $A(4p_{1/2})/g_I$ values for potassium isotopes when the $g$-factors were measured independently. Dots with error bars -- experimentally measured $A(4p_{1/2})$ constants from Refs.~\cite{k37p_exp,K_st_p,papuga14} divided by the nuclear $g$-factors from Refs.~\cite{k37_exp,39K_mu,stone19,k42_exp}. Dotted line -- weighted mean value for these isotopes.}
\end{figure}
\section{Evaluation of the nuclear magnetic moments}
Previously the nuclear $g$-factors of potassium isotopes far from stability were extracted from the $A(4s_{1/2})$ constants~\cite{papuga14} neglecting the RHFA. The additional uncertainties of 0.3\% and 0.5\% were added for odd-even and odd-odd isotopes, respectively, to account for RHFA. Note, that this estimation of the RHFA contribution is based on the experimental data with uncertainties $\sim50-100$\% and theoretical calculations with unknown accuracy, therefore, the conservative estimation of the additional uncertainties due to RHFA should be 0.5\% and 0.8\% in odd-even and odd-odd cases, respectively (see Table~2 in~\cite{papuga14}). However, the $A(4p_{1/2})$ constants are more convinient for the nuclear magnetic moments extraction. Due to negligible magnitudes of both BR and BW corrections the $A(4p_{1/2})/g_I$ values should be the same for different potassium isotopes. In order to estimate this value from experimental data we use independently measured HFS constants and nuclear $g$-factors of $\rm ^{37,\,39-42}K$ isotopes (see~\Fref{exp}). 

\hfill
\break
The weighted mean value ${\cal A}_{0}^\mathrm{mean} (4p_{1/2}) = 106.44(8)$~MHz was used to extract the nuclear $g$-factor: $g_I = \tfrac{A(4p_{1/2})}{{\cal A}_{0}^\mathrm{mean} (4p_{1/2})}$. The comparison of our results to the literature values from Ref.~\cite{papuga14} with uncertainties due to RHFA increased in accordance with more conservative prescription outlined above is presented in~\Tref{tbl:g_factors}. New results yield smaller uncertainties than the literature data~\cite{papuga14}, except $\rm ^{51}K$ due to large relative error of experimental HFS constant.
\begin{table}[tbh]
\caption{\label{tbl:g_factors}
The nuclear magnetic moments of potassium isotopes extracted from experimentally measured $A (4p_{1/2})$ constants~\cite{papuga14}.
The results are compared to the literature data.}
\begin{tabular}{ccrrr}
\hline
\\[-3mm]
\setlength{\tabcolsep}{25pt}
Isotope&$I^\pi$& $A (4p_{1/2})$, MHz&\multicolumn{2}{c}{$\mu$, $\mu_N$}\\
&&Ref.~\cite{papuga14}&\multicolumn{1}{c}{this work}&\multicolumn{1}{c}{Ref.~\cite{papuga14}} \\
\hline \\[-3mm]
38&$3^+$&48.9(2)&1.378(6)&1.371(12)\\
44&$2^-$&$-45.8(2)$&$-0.861(4)$&$-0.857(8)$\\
46&$2^-$&$-55.9(2)$&$-1.050(4)$&$-1.046(9)$\\
47&$1/2^+$&411.8(2)&1.934(2)&1.929(10)\\
48&$1^-$&$-96.3(3)$&$-0.905(3)$&$-0.900(7)$\\
49&$1/2^+$&285.6(7)&1.342(3)&1.339(7)\\
51&$3/2^+$&36.6(9)&0.516(13)&0.513(5)\\
\hline 
\end{tabular}
\end{table}
\hfill
\break
\begin{table}[tbh]
\caption{\label{tbl:d_nuc}
The $d_\mathrm{nuc}$ factors determinated from RHFA values~\cite{Per13,papuga14} by \Eref{rhfa2} with $^{39}\mathrm{K}$ as the reference isotope. For comparison the $d_\mathrm{nuc}$ factors calculated within the single-particle nuclear model are given in the last column.}
\begin{tabular}{crrcrrr}
\hline
\\[-3mm]
\multicolumn{2}{c}{Isotope\ \  $I^\pi$}&\multicolumn{2}{c}{$g$-factor}&\multicolumn{1}{c}{$\prescript{39}{}\Delta^{*}$, \%}&\multicolumn{2}{c}{$d_\mathrm{nuc}$}\\
&&&&\multicolumn{1}{c}{Eq.\eqref{rhfa}}&\multicolumn{1}{c}{Eq.\eqref{rhfa2}}&\multicolumn{1}{c}{Eq.\eqref{F2}} \\
\hline \\[-3mm]
37&$3/2^+$&$0.13547$(4)&\cite{k37_exp}&$-$0.249(35)&$-4.6$(4)&$-5.3$ \\
39&$3/2^+$&$0.2609775$(2)&\cite{39K_mu}&\multicolumn{1}{c}{0.0}&\multicolumn{1}{c}{$-$}&$-2.1$\\
40&$4^-$  &$0.324493$(8) &\cite{stone19}&0.466(19)&2.7(2)&$-$\\
41&$3/2^+$&$0.143248$(3)&\cite{stone19} &$-$0.22936(14)&$-4.4(1)$ & $-5.0$\\
42&$2^-$ &$-0.57125$(3)&\cite{k42_exp} &0.336(38)&1.3(4)&$-$\\
47&$1/2^+$ &3.869(3)&&0.272(90)&0.7(9)&1.0\\
\hline 
\end{tabular}
\end{table}
\section{Evaluation of nuclear factors}
For a number of potassium isotopes the relative hyperfine anomalies are known with sufficient accuracy~\cite{Per13}.
For the reference isotope, $^{39}$K, the single-particle nuclear model [\Eref{F2}] gives $d^{(39)}_\mathrm{nuc} = -2.1$.
This factor corresponds to BW correction $\varepsilon^{(39)} = -0.205\%$.

Nuclear factors for $\rm ^{37,\,41}K$ calculated by the single-particle nuclear model (see column 6 in \Tref{tbl:d_nuc}) happen to be lower than corresponding experimental values (see column 5 in \Tref{tbl:d_nuc}). Note, that magnetic moment of $\rm ^{39}K$ (0.39~$\mu_N$) is nearly twice as large as magnetic moments of $\rm ^{37,\,41}K$ ($\sim 0.20\,\,\mu_N$) with the same spin ($I^\pi=3/2^+$) and leading nuclear configuration $\pi d_{3/2}$. Correspondingly, $d_\mathrm{nuc}(^{37,\,41}\mathrm{K}) \approx 2\times d_\mathrm{nuc}(^{39}\mathrm{K})$ and single-particle evaluations underestimate $d_\mathrm{nuc}$ for $\rm ^{37,\,41}K$.
Keeping in mind the strong single-particle nature of the $\rm ^{39}K$ ground state, this disagriment indicates mixing of the nuclear configurations in $\rm ^{37,\,41}K$.

Surprisingly, similar jump-like behavior was found for gold nuclei with $I^\pi=3/2^+\,(\pi d_{3/2})$: \mbox{$\mu(^{199}\mathrm{Au})=0.27~\mu_N$} whereas $\mu(^{191,\,193,\,195,\,197}\mathrm{Au}) \approx 0.15~\mu_N$ and $d_\mathrm{nuc}(^{191,\ldots,197}\mathrm{Au}) \approx 2\times d_\mathrm{nuc}(^{199}\mathrm{Au})$~\cite{dem20}. Besides, the single-particle model does not describe well the $d_\mathrm{nuc}$ parameter in light Au isotopes~\cite{dem20}. Similarly, this model fails to reproduce the experimental $d_\mathrm{nuc}$ value for $\rm ^{37,\,41}K$. This similarity supports the assumption of Ref.~\cite{dem20} that in contrast to a rather pure ground state of $\rm ^{199}Au$, the ground state of $\rm ^{197}Au$ (and lighter odd Au isotopes with $I^\pi=3/2^+$) has a noticeable admixture of other configurations.
\break
\section{Conclusions}
We calculate the hyperfine structure constants of low-lying states of potassium atom taking into account the Bohr--Weisskopf and Breit--Rosenthal effects. In order to separate these effects we use two cases of nuclear magnetization distribution and the same homogeniouslly distributed charge. The first case describes the point-like magnetic dipole in the center of the nucleus, whereas the second assumes the homogeneously magnetized sphere of nuclear radius. We extract atomic parameters $b_N$, $b_M$, and ${\cal A}_0$ for each considered state. To estimate the BW correction we assume the atomic-nuclear factorization and use the $d_\mathrm{nuc}$ factor. 

According to our calculations the $4p_{1/2}$ state of K I is almost free from both BR and BW corrections. Using this fact, we obtain the mean value ${\cal A}_0^\mathrm{mean} (4p_{1/2}) = 106.41(8)$~MHz from experimental data. The result of our LCC calculations agrees with this value within 1.2\%. We use the ${\cal A}_0^\mathrm{mean}$ value to extract the nuclear magnetic moments of short-lived potassium isotopes from $A(4p_{1/2})$ constants.

Experimentally measured relative hyperfine anomalies provide the relation between the $d_\mathrm{nuc}$ factors of different isotopes. One can consider the configuration of $^{39}$K nuclear ground state as the single proton hole with respect to the doubly magic $^{40}$Ca. This justifies our choice to use the single-particle $d_\mathrm{nuc}^{(39)} = -2.1$ as a reference to restore the nuclear factors for other isotopes from RHFA values. The striking similarity of the jump-like behavior of magnetic moments and $d_\mathrm{nuc}$ parameters in K and Au isotopes supports the assumption of a configuration mixing in light odd Au isotopes with $I^{\pi} = 3/2^+$~\cite{dem20}.
\subsection*{Acknowledgments}
We thank Prof. M.S. Safronova for helpful discussions and providing the LCC code.
This research was funded by the Russian Science Foundation Grant \textnumero 20-62-46006.
\bibliographystyle{apsrev}

\end{document}